\documentclass[twocolumn]{aastex631}

\usepackage{natbib}

\shorttitle{X-ray Absorption Spectrum of RX J$0822-4300$}
\shortauthors{Groger et al.}
\begin{document}

\title{Evidence for Atomic Absorption Features in the High Resolution X-ray Spectrum of\\
 the Neutron Star in Puppis A}

\author[0000-0002-7054-9053]{John Groger}\altaffiliation{now in the Department of Physics, Washington University in St. Louis}
\affiliation{Department of Astronomy and Columbia Astrophysics Laboratory,
Columbia University\\
550 West 120th Street,
New York, NY 10027, USA}
\altaffiliation{now in the Department of Physics, Washington University in St. Louis}

\author[0000-0001-9225-6481]{Frits Paerels}
\affiliation{Department of Astronomy and Columbia Astrophysics Laboratory,
Columbia University\\
550 West 120th Street,
New York, NY 10027, USA}

\author[0000-0002-9870-2742]{Slavko Bogdanov}
\affiliation{Department of Astronomy and Columbia Astrophysics Laboratory,
Columbia University\\
550 West 120th Street,
New York, NY 10027, USA}

\author[0000-0003-3847-3957]{Eric V. Gotthelf}
\affiliation{Department of Astronomy and Columbia Astrophysics Laboratory,
Columbia University\\
550 West 120th Street,
New York, NY 10027, USA}

\author[0000-0002-6822-4823]{David J. Helfand}
\affiliation{Department of Astronomy and Columbia Astrophysics Laboratory,
Columbia University\\
550 West 120th Street,
New York, NY 10027, USA}

\author[0000-0001-8816-236X]{Ivan Hubeny}
\affiliation{Department of Astronomy and Steward Observatory \\
933 North Cherry Avenue,
Tucson, AZ 85721, USA}

\author{Thierry Lanz}
\affiliation{Observatoire de la Cote d’Azur, Nice, France}

\author[0000-0001-8748-5466]{Thomas A. Gomez}
\altaffiliation{George Ellery Hale Fellow}
\affiliation{Department of Astrophysical and Planetary Sciences, University of Colorado, Boulder, CO 80305, USA}
\affiliation{Laboratory for Atmospheric and Space Physics, University of Colorado Boulder, Boulder, CO 80303, USA}
\affiliation{National Solar Observatory, University of Colorado Boulder, Boulder, CO 80303, USA}
\affiliation{Department of Astronomy, University of Texas at Austin,
 Austin, TX 78712, USA}

\begin{abstract}

We present evidence for atomic absorption lines in the high-resolution $4-30$ \AA\ X-ray spectrum of the neutron star RX J$0822-4300$ in the supernova remnant Puppis A.
Comparison with model atmosphere calculations shows that features in the observed spectrum can be uniquely associated with redshifted and pressure-broadened transitions in highly ionized oxygen and neon. We also spectroscopically confirm the previously estimated strength of the surface magnetic dipole field; we detect both the linear and the quadratic Zeeman effect.
We derive values for both the gravitational redshift and the acceleration of gravity at the stellar surface, yielding the first purely spectroscopic estimates for the radius and mass of a neutron star.

\end{abstract}


\section{Introduction} \label{sec:intro}

We observed RX J$0822-4300$, the neutron star in the supernova remnant Puppis A, over the period March 2021-November 2021, for a total exposure of 385.24 ksec spread over 14 separate observations, using the Low Energy Transmission Grating Spectrometer (LETGS; \citep{brinkman2000} on the {\it Chandra} X-ray Observatory (CXO; \citealt{weisskopf2000}) recorded with the HRC-S focal plane detector. 
RX J$0822-4300$ is a bright X-ray point source \citep{Petre1996}, the remnant of a core-collapse supernova $\sim 3700$ years ago \citep{Winkler1988}. The supernova remnant shows ejecta enriched in O, Ne, Mg, Si, and Fe in optical and X-rays \citep{Winkler1985,CanizaresWinkler1981,Mayer2022}. The distance has recently been redetermined as $1.3 \pm 0.3$ kpc \citep{Reynoso2017}.

The point source is a confirmed thermal emitter \citep{Petre1996}, with a spectrum approximated by the sum of two blackbodies ($kT_{1,2} = 0.222,0.411 $ keV), associated with two hot spots on the surface, approximately in antiphase \citep{alford2022}. From the X-ray spin period 0.112 s and spindown rate $(9.28 \pm 0.36) \times 10^{-18}$ s s$^{-1}$, \citet{alford2022} determine a surface dipole magnetic field strength of $B = 2.9 \times 10^{10}$ G. 

Of the known Central Compact Object (CCO) X-ray sources (thermally emitting neutron stars located in young supernova remnants; \citealt{DeLuca2017}), the spectra of the neutron stars 1E $1207.4-5209$ and the CCO in Cassiopeia A have been interpreted as showing evidence for photospheric atomic absorption. In the case of 1E$1207.4-5209$ absorption by mid-$Z$ elements has been suggested \citet{HaileyMori2002}, though a series of electron cyclotron resonances in a field of $B = 8 \times 10^{10}$ G is now considered the correct interpretation, since the spectroscopic and spindown values for the magnetic field agree \citep{Gotthelf2020}. The $E > 1$ keV X-ray flux of the CCO in Cas A compared to the extrapolation of a blackbody normalized to a reasonable fraction of the expected neutron star surface area (no significant pulsations are detected) suggests photospheric absorption, possibly by highly ionized carbon \citep{HoHeinke2009}. None of these inferences are based on direct spectroscopic identification, however. 

Should the photosphere of a neutron star contain heavy elements, then the absorption spectra of oxygen, neon, magnesium, and silicon should be detectable with the grating spectometers on {\it Chandra} for stars with effective temperatures of order a few million K. The CCO's are located in regions of significant diffuse emission (their SNR), but the angular resolution of {\it Chandra} produces narrow, high contrast spectral images for the faint point source against the nebular background. Interstellar absorption towards the CCO's limits observations to photon energies above about 400 eV in practice, so a spectroscopic search for carbon is unfortunately impossible. Of all the CCO's, RX J$0822-4300$ has the most favorable combination of low absorption and brightness. 

In the following, we will briefly describe the observation and data analysis. We then review some atomic spectroscopy relevant to the interpretation of our spectrum (Section \ref{sec:Zeemanetal}). We present our model atmospheres code,
{\sc TLUSTY}, adapted to the modeling of neutron star atmospheres, in Section \ref{sec:modelatmos}. We describe recent detailed line-broadening calculations that have been implemented in {\sc TLUSTY} for this work, and we present a few characteristic model atmosphere spectra to establish the salient spectral features we expect to see. We then present the observed spectrum and our spectroscopic interpretation, including evidence for atomic absorption features due to highly ionized oxygen and neon. We derive bounds on the stellar parameters from the spectrum, in particular on the gravitational redshift and the acceleration of gravity at the surface, and estimate the implied values for the stellar mass and radius (Section \ref{sec:conclusion}).

\section{Observation} \label{sec:observation}
We obtained a 385.24 ksec {\it Chandra} LETGS/HRC-S observation of RX J$0822-4300$. The exposure was spread over 14 observations conducted between March 25 and November 29, 2021, with 14 roll angles roughly evenly spread between 291 and 65 degrees. The detailed data analysis, and a thorough empirical analysis of the spectrum will be fully described in a companion paper \citep{Gotthelf2024}.

Initial extraction of the grating spectrum was performed using {\tt CIAO} version 4.13.\footnote{\tt CIAO: \url{https://cxc.cfa.harvard.edu/ciao/}.} We estimated and subtracted background photons due to the remnant using a source extraction box of cross-dispersion width 55 pixels, and background boxes taken above and below the dispersed light from the point source. We then combined source and background spectra from the 14 separate exposures. In a Python script, we fit a fifth-order polynomial to the smooth background and binned the data to bins of width 0.1 and 0.2 \AA. The background is smooth, and the coefficients of the polynomial are well-determined. We therefore assume that the expected level of counting statistics fluctuations in the background-subtracted spectrum is dominated by the Poisson fluctuations in the raw (before background subtraction) spectrum. Finally, we produced histograms of background-subtracted photon counts, which show evidence for spectral features in the range from approximately 3-30 \AA. Because of the customized nature of our models (Section \ref{sec:modelatmos}), our data analysis centered on these histograms rather than standard spectral fitting in {\tt xspec}. All models were multiplied with the HRC-S effective area data for Cycle 21 
\footnote{{\tt \url{https://cxc.harvard.edu/cgi-bin/prop\_viewer/build\_viewer.cgi?ea}}}.

\section{Zeeman, Stark, and Einstein} \label{sec:Zeemanetal}

We briefly review some aspects of atomic spectroscopy at high density and high magnetic field strength in order to aid the interpretation of the observed spectrum of RX J$0822-4300$ and our model atmospheres spectra. We expect the effective temperature of the star to be several times $10^6$ K, so the constituent elements of the atmosphere are expected to be highly ionized. In particular, the low- and mid-$Z$ elements are expected to be in their H- and He-like charge states. Our spectrum covers the range of the H- and He-like spectra of the elements O through Si. The density is expected to span the range $n_e = 10^{21-25}$ cm$^{-3}$ for surface acceleration of gravity $\log g = 14.3-14.6$, and temperatures of order a few million K. Gotthelf, Halpern, \& Alford \citep{gotthelf2013}, from spindown measurements, have determined a surface dipole magnetic field strength $B = 2.9 \times 10^{10}$ Gauss. 
For reference, we will take the \ion{O}{8} Ly$\alpha$ transition, involving energy levels at $-871.41$ eV ($1s$), $-217.91$ eV ($2s$), and $-217.92,-217.73$ eV ($2p_{1/2,3/2}$) in the absence of external perturbations. We will estimate the effects of the various perturbations on the bound energy levels. Note that these estimates are very rough, and are just meant to illustrate the scale of the effects. Our quantitative analysis is of course based on precise calculations.  

%{\color{red} TG:

Atomic structure and spectra are significantly modified in the presence of a strong magnetic field. 
At low magnetic fields, energy levels are split according to the atom's $m$ quantum number, at higher magnetic field, the energy levels form the classic triplet pattern, known as the Paschen-Back limit \citep{Cowan81}.
Beyond this, the quadratic Zeeman effect will cause the ionization potential of the lower-energy states to increase \citep{Ruder94}, blueshifting the atomic transitions.
The strength of the magnetic interactions can be characterized by comparing the radius of the first Bohr orbit in a hydrogenic ion of nuclear charge $Z$, to the Larmor radius: they are equal at
$B = Z^2\ m_e^2 c (e/\hbar)^3 = 2.35 Z^2 \times 10^9$ G.
In a field of $B \approx 3 \times 10^{10}$ Gauss, \ion{O}{8} is still dominated by the Coulomb field of the nucleus.
The characteristic energy scale for the linear Zeeman effect is $\Delta E = \mu_{\rm B} B$, with $\mu_{\rm B}$ the Bohr magneton and $B$ the field strength. In our case, we get
$\Delta E = 174\ (B/3 \times 10^{10}\ {\rm G})$ eV. 
The quadratic Zeeman shift originates from the diamagnetic term in the Hamiltonian $H_{\rm DM} = e^2B^2 r^2/(8m_e c^2)$ (Gaussian units), with $e$ the electron charge, $m_e$ the electron mass, $r$ the position of the electron with respect to the nucleus. 
Using Bohr orbits ($r = n^2 a_0/Z$ with $n$ the principal quantum number, $a_0$ the Bohr radius, and $Z$ the nuclear charge), we get ($Z=8$) $\Delta E = 7.73\ n^4 (B/3 \times 10^{10}\ {\rm G})$ eV. For \ion{O}{8} Ly$\alpha$, this implies a net blueshift of 116 eV.

In the dense neutron star atmosphere, the spectra are also altered by the plasma environment. 
The result is that the spectral lines will broaden and the ionization threshold will be lowered. Due to the disparate timescales of their motions, electrons are treated within a collision broadening formalism, while ions are often treated as static \citep{Griem74}.
At the magnetic fields under discussion, H-like systems have lost their degeneracy and are in the isolated-line limit.
In this isolated line limit, ions contribute very little to the line width \citep{Alexiou14}, and electron broadening dominates.
%In the impact limit, the widths of isolated spectral lines due to electron collisions are given in terms of collision amplitudes and cross sections \citep{Baranger},
%\begin{equation}
%    2w = \Bigg\{n_ev\left[\sigma_{in~u} + \sigma_{in~l} + \int d\Omega |f_u(\Omega)-f_l(\Omega)|^2\right]\Bigg\}_{Av}
%\end{equation}
%where $n_e$ and $v$ are the electron density and velocity, respectively, $\sigma_{in}$ are inelastic cross sections, $f$ are the collision amplitudes, the subscripts $u$ and $l$ designate the upper and lower state, respectively, of the transition, and the average is over all directions $\Omega$.
The collisional width can be evaluated very roughly by approximating the collision cross section with the Weisskopf radius, $\sigma \approx \pi \rho_{\rm W}^2$, where $\rho_{\rm W}$ is the Weisskopf radius.
Even this rough estimate shows that electron collision is the most important broadening mechanism even without accounting for the magnetic field, with
$\Delta E \sim 30 Z^2 (n_e/10^{23}\ {\rm cm}^{-3}) (T/10^6\ {\rm K})^{-1/2} $ eV; the Weisskopf radius has been taken from \citet{Oks2018}.
However, a detailed fully-quantum mechanical electron broadening model that includes magnetic field is currently being developed \citep{Gomez2023}.
This model accounts for the Landau quantization of the free electrons.
In our case, the cyclotron energy (Landau level spacing) is about 350 eV, while the thermal energy of the electrons is $260 (T_e/3 \times 10^6\ {\rm K})$ eV, so there will be a significant population in the first few excited Landau states.
These detailed models have found that exchange interactions (which account for the indistinguishability of electrons) dominate the collision cross sections.
For \ion{O}{8} at these magnetic fields, the electron broadening far exceeds other broadening mechanisms, such as ion Stark and motional Stark broadening \citep{Gomez2024b}.

For completeness, the measured 0.112 s spin period of the star implies a rotational Doppler broadening of only $\Delta\lambda/\lambda \leq 1.87 \times 10^{-3} (R/10\ {\rm km})$.  

Finally, we note and emphasize that the quadratic Zeeman effect causes a net blueshift of all transitions, opposite to the gravitational redshift.

\begin{figure*}
    \centering
    \includegraphics[scale=0.6]{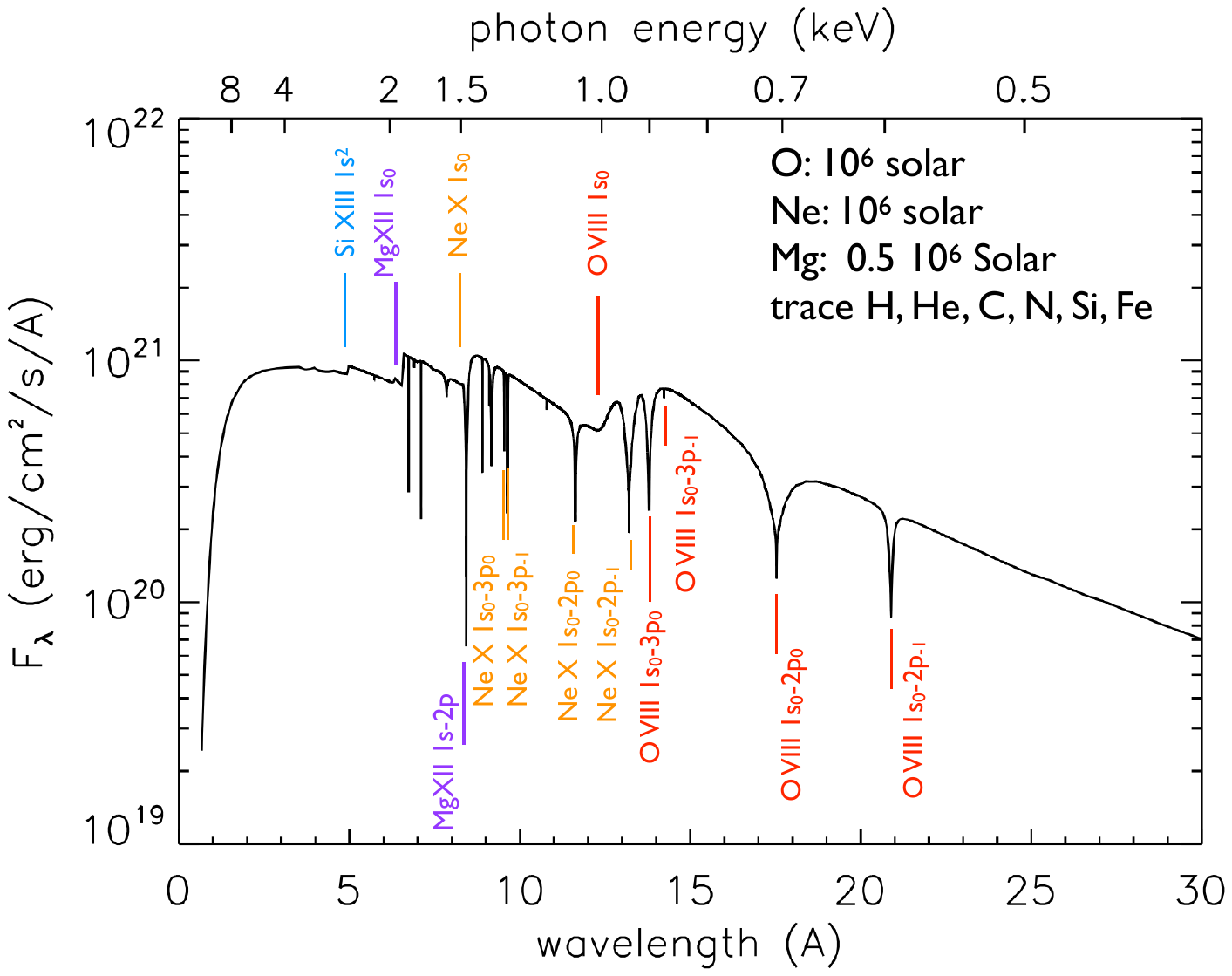}
    \vskip -0.4 in
    \caption{Model atmosphere spectrum for a hot neutron star with an oxygen-neon dominated atmosphere. The effective temperature is $T_{\rm eff} = 4.6 \times 10^6$ K, $\log g = 14.6$, $B = 2.9 \times 10^{10}$ Gauss. The most prominent atomic features have been labeled. The wavelength scale is the rest frame of the neutron star (no redshift). A nomogram for the corresponding photon energies is displayed at the top; a simple conversion is $E [{\rm keV}] =
    12.3984/\lambda[{\mathrm\AA}]$.
    }
    \label{fig:1}
\end{figure*}

\begin{figure*}
    \centering
    \includegraphics[scale=0.6]{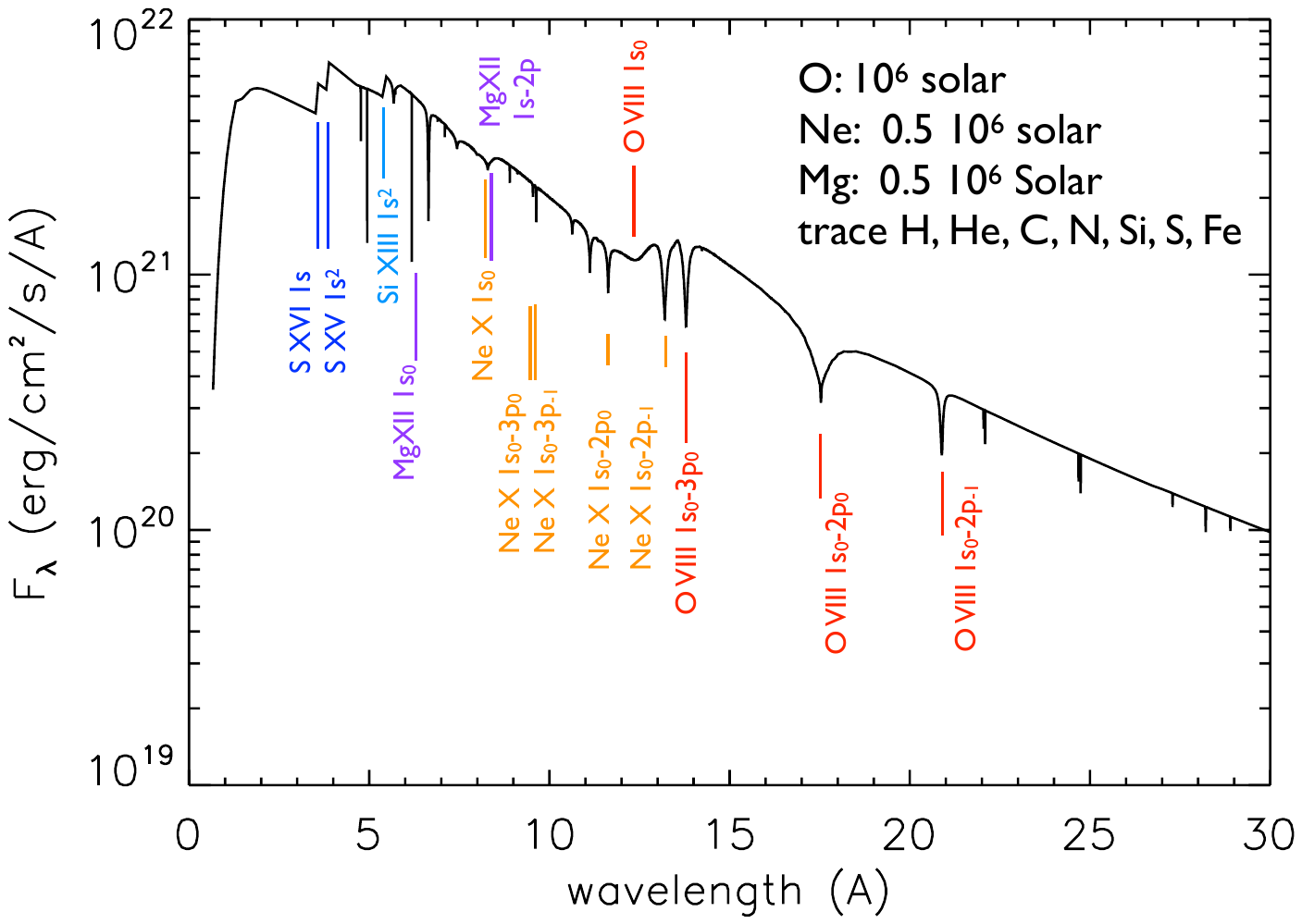}
    \vskip -0.5 in
    \caption{As \hyperref[fig:2]{Figure 1}, but for a higher temperature, $T_{\rm eff} = 5.9 \times 10^6$ K, and slightly different composition; note the S XV and XVI edges at 3.5 \AA. The wavelength scale is the rest frame of the neutron star (no redshift).
    }
    \label{fig:2}
\end{figure*}

\begin{figure*}
    \centering
    \includegraphics[scale=0.55]{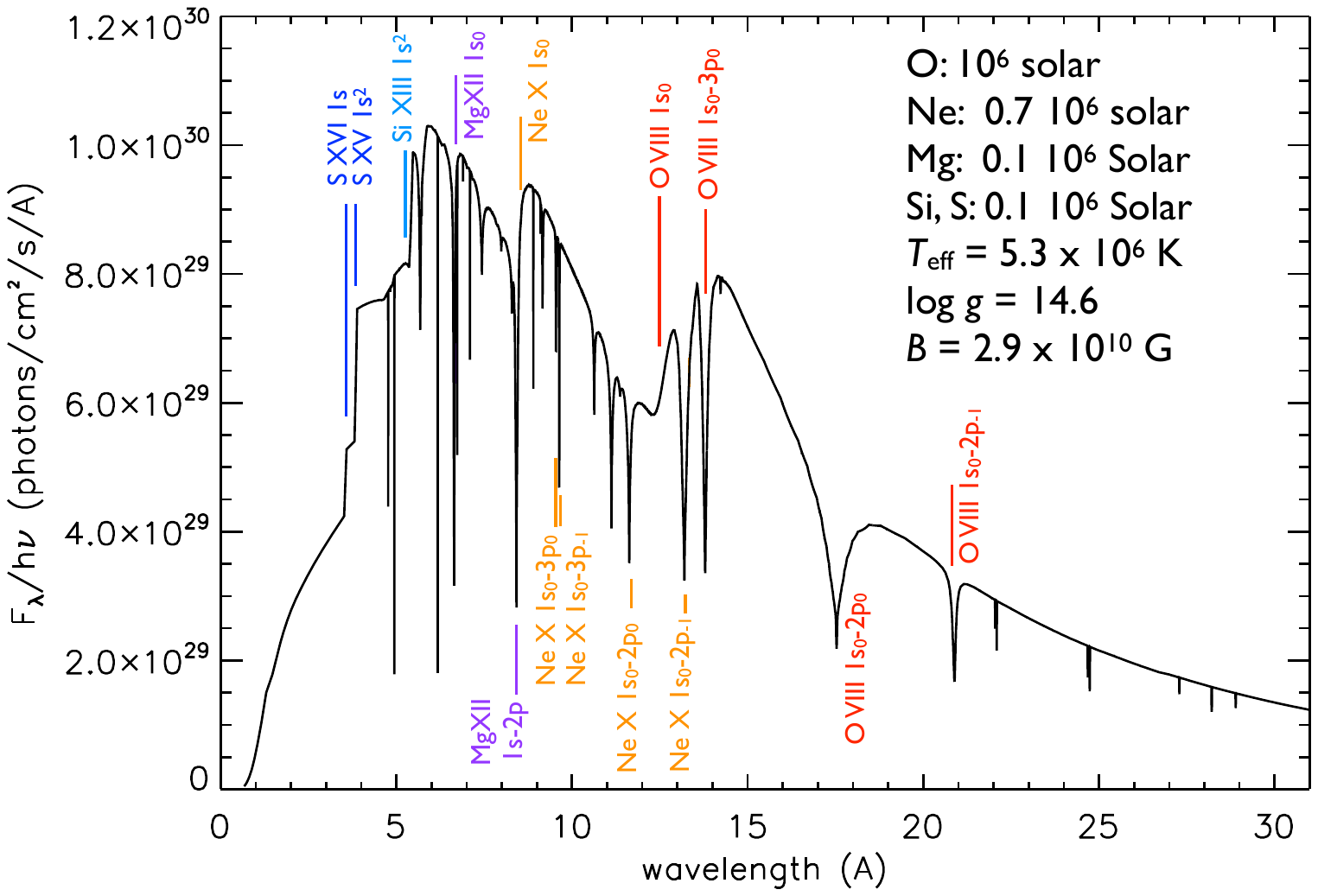}
    \vskip -0.5 in
    \caption{Model atmosphere spectrum at $T_{\rm eff} = 5.3 \times 10^6$ K, $\log g = 14.6$, $B = 2.9 \times 10^{10}$ Gauss, plotted on a linear flux scale. Note that this figure shows the photon flux, which is more relevant for a comparison to an observed photon count spectrum. The wavelength scale is the rest frame of the neutron star (no redshift).
    }
    \label{fig:3}
\end{figure*}

\section{Model Atmospheres Calculations} \label{sec:modelatmos}

We assume a plane-parallel, horizontally-homogeneous atmosphere in radiative and hydrostatic equilibrium. Model atmospheres were computed using the program {\sc TLUSTY} \citep{hubeny1995,hubeny2017}, appropriately modified for conditions in neutron star atmospheres.
Details of these calculations will be published elsewhere.

The densities are high enough that for the low- and mid-$Z$ elements, LTE is expected to apply. 
The effects of pressure ionization are treated by means of the occupation probability formalism of \citet{Hummer1988}, adopted for stellar atmosphere calculations by \citet{hubeny1994}. 
We assume a composition dominated by O and Ne, with traces of the other elements. The atomic structure of the H- and He-like ions of O and Ne in a strong magnetic field is calculated explicitly, and the lowest electric dipole-allowed members of each spectroscopic series are kept explicitly in the radiative transfer calculation; in practice, we kept transitions $n=1-2$ and $n=1-3$. In H- and He-like O, we approximated the effect of overlapping and merging high-order series members by adding a 'pseudocontinuum' opacity up to the series limit.

The collision broadening by free electrons
is included, with an explicit absorption profile that can be well approximated by a Lorentzian with a width that scales approximately proportional to density \citep{Gomez2023}.

We explored a range of effective temperatures $T_{\rm eff} \sim 2 \times 10^6 - 6 \times 10^6$ K and gravitational accelerations $\log g = 14.3 - 14.6$. The chemical composition is dominated by O and Ne, to which we add smaller amounts of Mg, Si, S, and Fe. H, He, C, and N are present in trace amounts. We show a few representative models to point out important spectroscopic features and their dependence on the physical variables. All models are shown on the 'rest-frame' wavelength scale, {\it i.e.} without a gravitational redshift.

In \hyperref[fig:1]{Figures 1 and 2} we show two models for $T_{\rm eff} = 4.6$ and $5.9 \times 10^6$ K, $\log g = 14.6$, with mostly O and Ne; the two models have slightly different chemical composition. The abundances are given in units of the Solar abundance (taken from \citep{Grevesse1998}), so the atmospheres are mostly pure O; for Oxygen, Neon, and Magnesium in their Solar number ratio, Ne/O = 0.18 and Mg/O = 0.056.
The surface magnetic field strength is $B = 2.9 \times 10^{10}$ G. Absorption lines and edges can be seen from H-like O and H- and He-like Ne, Mg, and Si. In the hotter model, we also added a trace of S. The magnetic field splits the resonance lines $1s-np$ into 
$1s_0-np_m$, with $m$ the component of the orbital angular momentum along the magnetic field direction. In a strong magnetic field, no spin flip transitions can occur in the dipole approximation, so $\Delta s = 0$ ($s$ is the spin quantum number), and we only have $\Delta l = 0,1$ for all significant absorption lines.

The dramatic pressure broadening on the O and Ne lines is obvious. Over a wide range of effective temperature (at least $T_{\rm eff} \approx 1-6 \times 10^6$ K) the peculiar shape of the spectrum in the $12-20$ \AA\ range remains qualitatively the same: the continuum between the pressure-broadened \ion{O}{8} $1s_0-2p_{0}$ line and the series limit at $\approx 12$ \AA\ almost looks like a very broad emission feature. The appearance of this stretch of continuum between effectively \ion{O}{8} Ly$\alpha$ and $\beta$ does not change much with magnetic field strength either, except that the wavelengths shift (and therefore the inferred gravitational redshift changes). The longest wavelength member of the \ion{Ne}{10} $n=1-2$ spectrum, $1s_0-2p_{-1}$, falls in the \ion{O}{8} series, for $B = 2.9 \times 10^{10}$ G. There is a significant \ion{Ne}{10} continuum edge at $\approx 8$ \AA. The short-wavelength continuum level and the shape of the Wien-like tail are of course strongly dependent on the effective temperature, but they are also sensitive to the abundances of Ne, Mg, Si, and S, through their blanketing effect. \hyperref[fig:2]{Figure 2} confirms these statements at higher effective temperature. 

\hyperref[fig:3]{Figure 3} shows a $T_{\rm eff} = 5.3 \times 10^6$ K model on a linear flux scale, a more realistic representation of the features. Again, the most prominent feature in the spectrum is the peculiar continuum between \ion{O}{8} Ly$\alpha$ and $\beta$ (12 to 18 \AA).

\newpage

\section{Analysis} \label{sec:analysis}

\begin{figure*}
    \centering
    \includegraphics[scale=0.64]{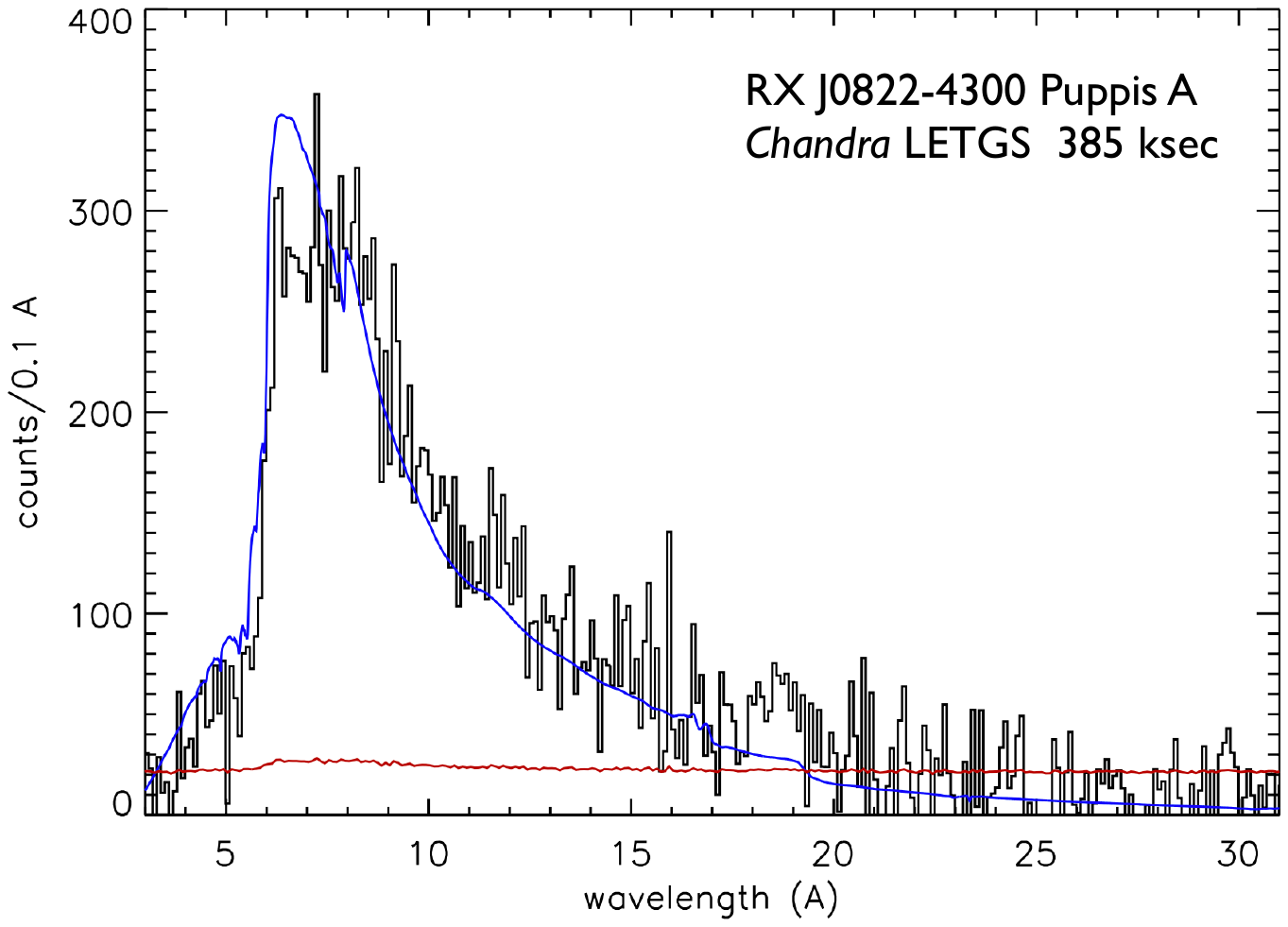}
    \vskip -0.6in
    \caption{Observed spectrum of RX J$0822-4300$, background-subtracted, binned in 0.1 \AA\ bins, as a function of wavelength (black histogram). The expected Poisson fluctuations in the spectrum are shown as the red curve. The blue curve is a model composed of two blackbodies at $kT_{\rm BB} = 4.69 \times 10^6$ K and $2.53 \times 10^6$ K (as seen by a distant observer), in the ratio 1.00:0.12, absorbed by $N_{\rm H} = 5.8 \times 10^{21}$ cm$^{-2}$. The wavelength scale is {\it Chandra}'s wavelength scale (so models have been redshifted).
    }
    \label{fig:4}
\end{figure*}

Figure \ref{fig:4} shows the spectrum of RX J$0822-4300$ observed with {\it Chandra} LETGS (both positive and negative spectral orders combined). The spectrum has been binned in 0.1 \AA\ bins (approximately two LETGS resolution elements), and the background (a nearly constant, approximately 450 counts/bin) has been subtracted. In red, we show the expected standard deviation in each bin, computed from Poissonian statistics. In blue, we show a model composed of two blackbodies, inspired by the analysis of previous CCD data \citep{alford2022}. The temperatures are $T_{\rm BB,hot} = 4.69 \times 10^6$ K and $T_{\rm BB,warm} = 2.53 \times 10^6$ K (as seen by a distant observer), and the relative normalizations are in the ratio $L_{\rm hot}/L_{\rm warm} = 1.0:0.12$. The luminosity ratio found by \citet{alford2022}, $L_{\rm hot}/L_{\rm warm} = 1.00:0.95$, overpredicts the flux in the 10-20 \AA\ range by $\sim 20-30\%$, when the flux is normalized at the peak near 7 \AA. The model has been attenuated with the transmission of cold interstellar gas for a column density of $N_{\rm H} = 5.8 \times 10^{21}$ atoms cm$^{-2}$, as derived for this model from the CCD data. The model has been multiplied with the LETG/HRC-S effective area for Observing Cycle 21; the steep edge at $\sim 6$ \AA\ is entirely due to the Au-M (gratings) and Ir-M (mirrors) absorption edges in the instrument effective area. To the first order spectrum, we added the predicted second and third order spectra. Longward of about 20 \AA, the higher orders rise and start contributing approximately half the detected counts (see below), dominated by third order. The orders $|m| > 3$ contribute very little to the measured spectrum, and only longward of 25 \AA.
This model has $\chi^2 = 290/159$, when evaluated over the range 4-20 \AA\ (the model with the normalization as in \citet{alford2022} has $\chi^2 = 304/159$). Additional structure is clearly visible in the spectrum, that cannot be accommodated by pure blackbody radiation. Note the mismatch between $\sim 6$ and 7 \AA. Varying the temperature of the hotter blackbody to bring down the flux in this band also eliminates the flux in the $3-5$ \AA\ band, making a serious mismatch in that range. Tuning a cooler blackbody to peak in the $7-10$ \AA\ range instead, raises the flux in the entire $7-20$ \AA\ range to unacceptably high levels. The blackbodies are simply too 'broad-band'.

\begin{figure*}
    \centering
    \includegraphics[scale=0.45]{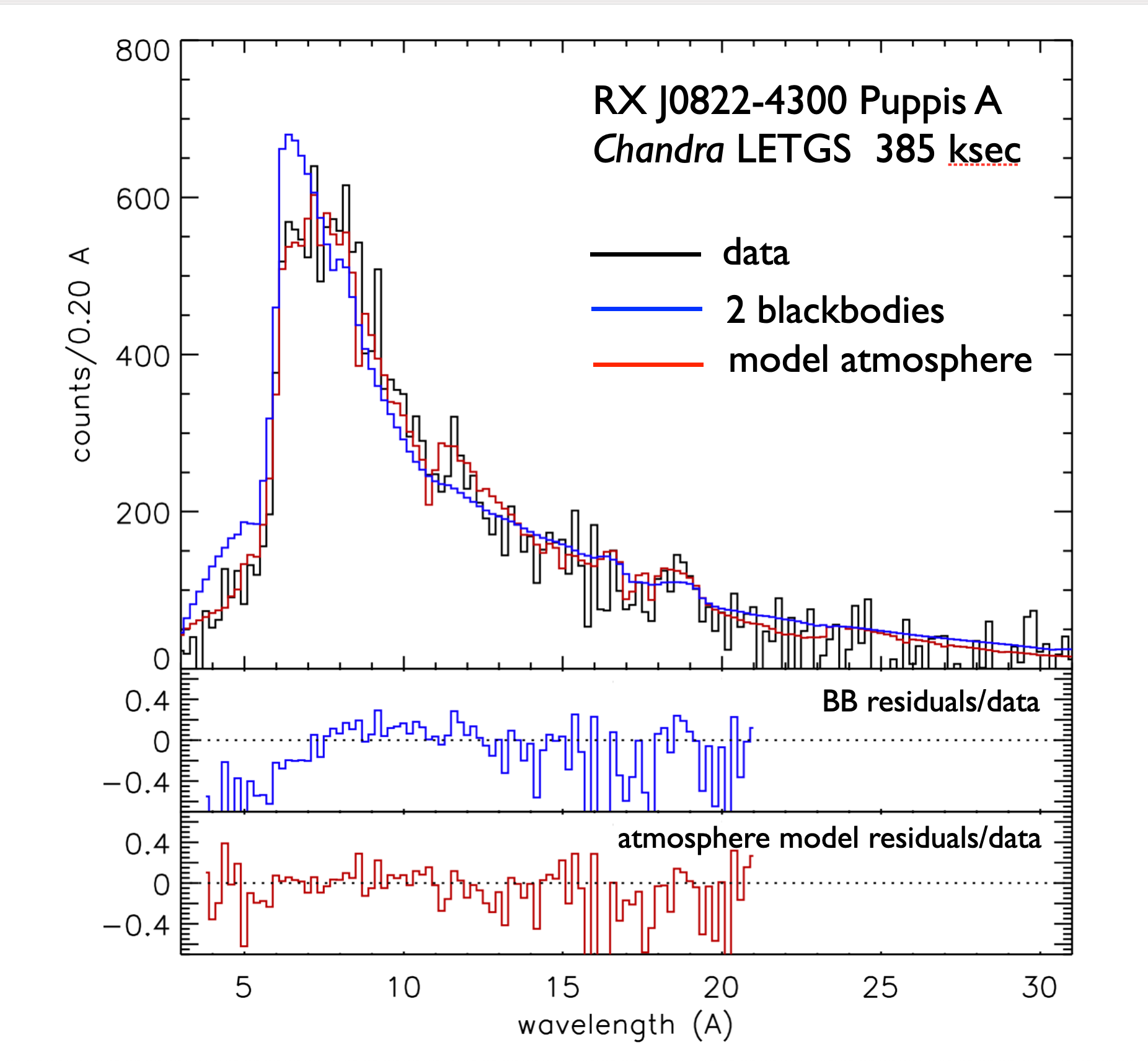}
    \vskip -0.0in
    \caption{
    Observed spectrum of RX J$0822-4300$, background-subtracted, binned in 0.2 \AA\ bins, as a function of wavelength (black histogram). The blue curve is a model composed of two blackbodies at $kT_{\rm BB} = 4.69 \times 10^6$ K and $2.53 \times 10^6$ K (as seen by a distant observer), in the ratio 1.00:0.12, absorbed by $N_{\rm H} = 5.8 \times 10^{21}$ cm$^{-2}$. The lower panel shows the residuals to the 2-blackbody model, expressed as the ratio (data$-$model)/data. We suppressed the parts of the residual spectrum entirely dominated by noise fluctuations, outside the range $3.7-21$ \AA. The mismatch between 3.7 and 7.0 \AA\ stands out. A depression between 15 and 17 \AA\ appears, as well as a positive feature between $\sim 17$ and $20$ \AA. The 2-blackbody model has $\chi^2 = 209$ for 85 bins of 0.2 \AA, in the $4-21$ \AA\ range. The red histogram is a model atmosphere spectrum, with $\chi^2 = 107$ for 85 bins of 0.2 \AA\ over the same range; residuals to this model are shown in the lowest panel, analogous to the residuals to the 2-blackbody model.
    }
    \label{fig:5}
\end{figure*}

To emphasize our point, that there is residual structure in the spectrum compared to a superposition of blackbodies, we show the residuals to the 2-blackbody fit in
Figure \ref{fig:5}, rebinned by a factor 2 to reduce the level of statistical fluctuations. In the lower panel, we show the residuals to the 2-blackbody fit, normalized to the data; we suppressed plotting the parts of the residual spectrum entirely dominated by noise fluctuations, outside the range $3.7-21$ \AA. The mismatch between 3.7 and 7.0 \AA\ stands out. A depression between 15 and 17 \AA\ appears, as well as a positive feature between $\sim 17$ and $20$ \AA. Also superimposed on the observed spectrum is a model atmosphere spectrum (see below). The values of $\chi^2$ on the 85 bins of 0.2 \AA\ in the $4-21$ \AA\ range are $\chi^2 = 209$ for the 2-blackbody model, and $\chi^2 = 107$ for the model atmosphere, a significant improvement. Residuals to the atmosphere model are shown in the lowest panel Figure \ref{fig:5}. To check whether either model is consistent with the measured spectrum, we also ran an Anderson-Darling test, testing the residuals against a normal distribution. The test fails to reject either model ($p = 0.218$ for the residuals with respect to the 2-blackbody model, and $p = 0.122$ for the atmosphere model, for 85 bins; a value $p \leq 0.05$ is assumed for rejection).
We therefore pursue the implications of using a model atmosphere to interpret the spectrum.

%HERE WE GO!!

\begin{figure*}
    \centering
    \includegraphics[scale=0.5]{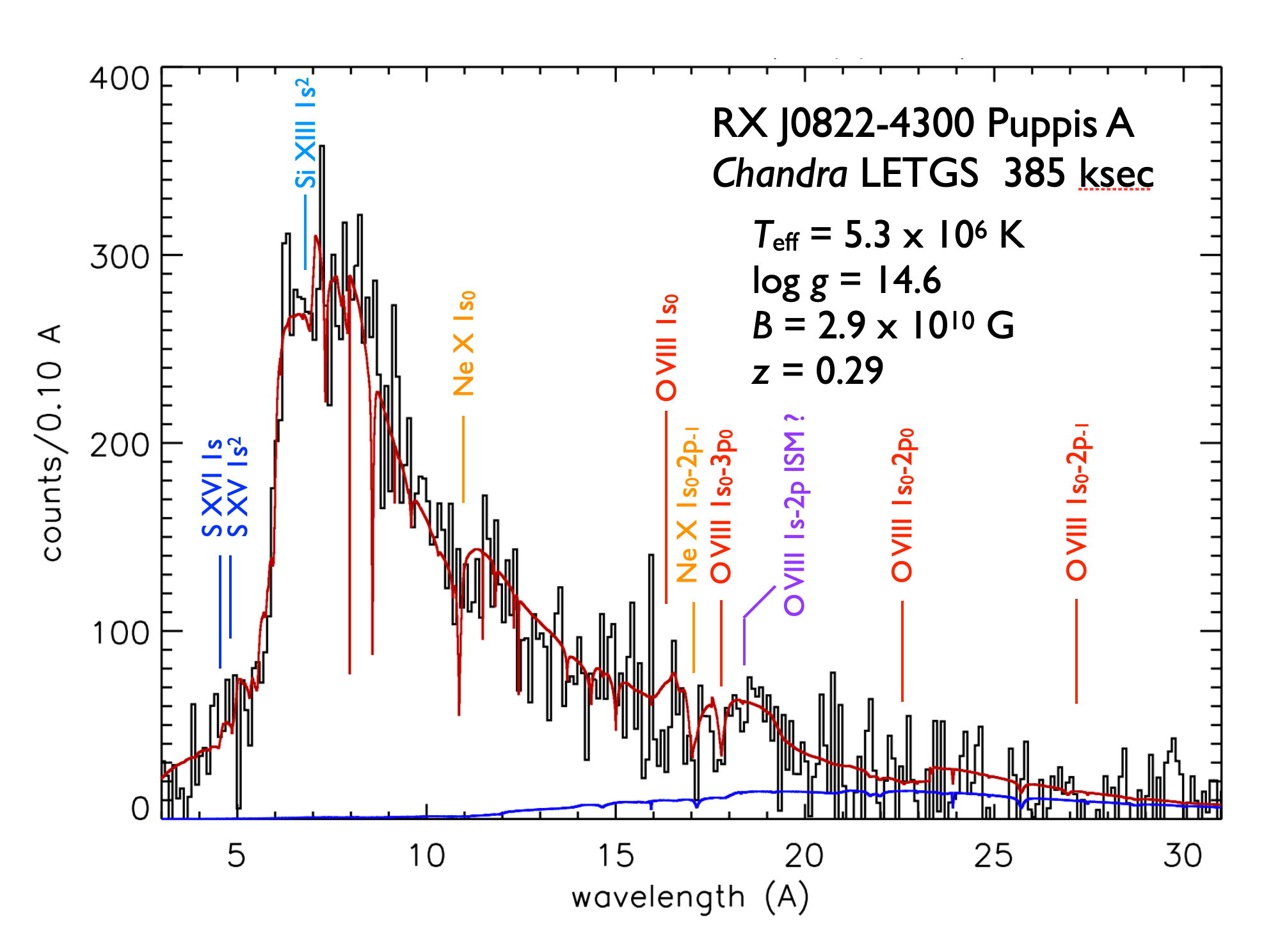}
    \caption{Observed spectrum of RX J$0822-4300$, background-subtracted, binned in 0.1 \AA\ bins, as a function of wavelength (black histogram). The red curve is a model atmosphere for $T_{\rm eff} = 5.3 \times 10^6$ K, $\log g =14.6$, $B = 2.9 \times 10^{10}$ G, absorbed by $N_{\rm H} = 2.2 \times 10^{21}$ cm$^{-2}$, and multiplied with the LETGS effective area; orders $|m| = 1,2,3$ have been added. The blue curve shows the sum of the second- and third spectral orders separated from the total. The wavelength scale is {\it Chandra}'s wavelength scale (so models have been redshifted). The location of photospheric absorption lines in O, Ne, Si, and S has been marked. The location of a possible absorption line produced in the hot gas of the supernova remnant has been marked in purple. 
    }
    \label{fig:6}
\end{figure*}

Figure \ref{fig:6} shows the observed spectrum, with a stellar model atmosphere superimposed, with model parameters as in \hyperref[fig:3]{Figure 3}: $T_{\rm eff} =  5.3 \times 10^6$ K, $\log g = 14.6$, $B = 2.9 \times 10^{10}$ G, and abundances (in units of the Solar abundances): O: $10^6$, Ne: $0.7 \times 10^6$, Mg: $0.1 \times 10^6$, Si and S: $0.1 \times 10^6$. The interstellar neutral H column density is $N_{\rm H} = 2.2 \times 10^{21}  $ cm$^{-2}$. 
The gravitational redshift has been set to $z = 0.29$. The model spectrum, shown in red, has been multiplied with the interstellar absorption and the LETGS effective area. The second and third spectral orders have been added; their contribution is shown explicitly, in blue.
The model has not been convolved with the spectrometer response (the 0.1 \AA\ bins are twice the width of the approximately 0.05 \AA\ FWHM Gaussian response). 
%With an optimized overall normalization, this model has $\chi^2 = 208/159$, calculated over the same spectral region as the two-blackbody model above. Between the blackbody model and the model atmosphere, the drop in $\chi^2$ is $\Delta\chi^2 = 82$. Though neither value of $\chi^2$ is the result of a formal fitting procedure, evaluating model parameters against a realization of the model in the presence of random noise (the data), the drop in $\chi^2$ does confirm the conclusion that the model atmosphere fits much better than the blackbody model. 

We draw attention to the $17.5-20$ \AA\ band: the bump seen in the observed spectrum coincides with the continuum between the strongest pressure-broadened \ion{O}{8} Ly$\alpha$ line and the Lyman series limit. Two strong absorption lines appear at $\sim 17$ and 18 \AA: the strongest \ion{O}{8} Ly$\beta$ line, and a \ion{Ne}{10} Ly$\alpha$ line. A step-like feature at $\sim 11$ \AA\ is the \ion{Ne}{10} Lyman series limit. At $\sim 7$ \AA, the He-like Si absorption edge cuts down the continuum by about 20\%, eliminating the clear mismatch seen in the two-blackbody model. An atmosphere model without additional opacity in the $3-5$ \AA\ band overpredicts the flux significantly; the presence of H- and He-like S absorption edges provides the necessary opacity. We note that the equivalent widths of any observed features are in fact in the range of predictions from pressure broadening. None of these features would ever be observable if it were not for that broadening mechanism.

As can be seen, most of the dispersed radiation longward of $\sim 21$ \AA\ is real, but dominated by higher spectral orders; the first order flux is suppressed by interstellar absorption.

The model we present here is not the result of a formal fitting procedure, which would likely not be informative, given the limited signal-to-noise in the observed spectrum. Instead, we match up prominent, robust, and stable features in the model spectra with features in the data that the two-blackbody model cannot reproduce. We started from the measured surface value of the magnetic dipole field 
$B = 2.9 \times 10^{10}$ G, and inspected the variation in wavelength of the strongest absorption lines and edges as function of field strength. Note that
it is important to remember that the dipole field strength derived from $P,\dot{P}$
is
quoted for an assumed stellar radius (10 km), that this field will have a
distribution along the stellar surface (but with the highest fields located in
the hottest parts of the atmosphere, if heat indeed flows along the B-field
from the core), and that higher multipole moments may contribute to the
local field in the stellar photosphere (which is what the ions in the
atmosphere see). In this context, we do note that the CCO 1E$1207.4-5209$ has shown glitches, which may be related to evolution of higher-order multipole moments of the magnetic field, present near the stellar surface. Still, the strength of the dipole component derived from the spin period evolution of this star agrees with the spectroscopically determined field strength at the photosphere
\citep{Gotthelf2020}, and so we start from the same assumption for our object.

\begin{figure*}
    \centering
    \includegraphics[scale=0.7]{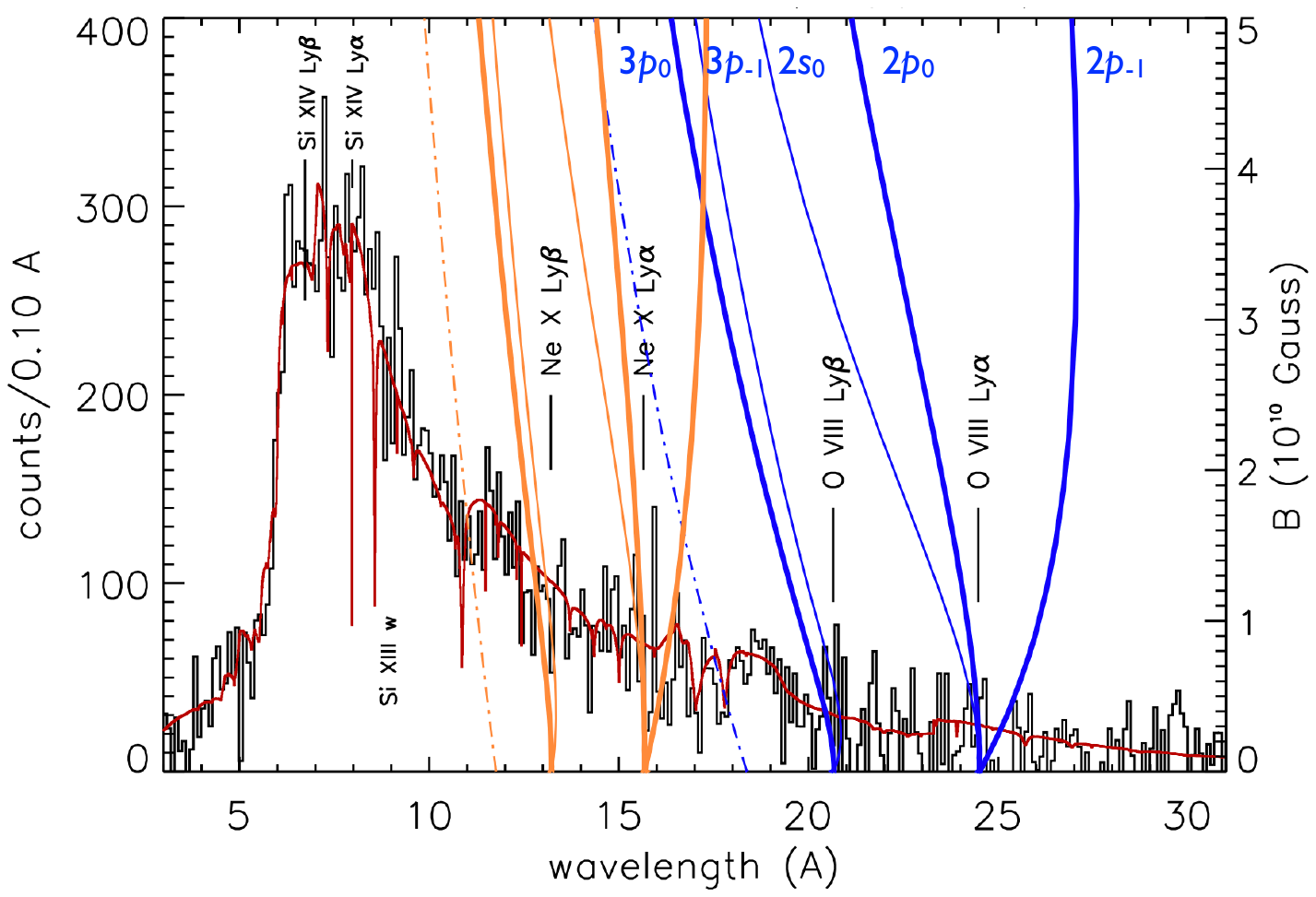}
    \vskip -0.7in
    \caption{Same as \hyperref[fig:6]{Figure 6}, with the magnetic field strength dependence of the strongest transitions in oxygen and neon superimposed. The right hand scale shows the field strength. Labels indicate the zero-field wavelengths of \ion{O}{8} and \ion{Ne}{10} Ly$\alpha$ and $\beta$. The transitions shown are, for both oxygen and neon, in blue and orange, respectively, from the longest to the shortest wavelength, $1s_0-2p_{-1}, 1s_0-2p_0,1s_0-2s_0; 1s_0-3p_{-1}, 1s_0-3p_0$. The $1s_0$ series limit is indicated by a dash-dot line. Weaker transitions are plotted with thinner lines. The upper levels of the transitions have been labeled for oxygen; the pattern is the same for neon.
    At low magnetic fields, we see the characteristic split $\Delta E = \mu_{\rm B} B$ due to the Paschen-Back effect; at higher fields, the quadratic Zeeman effect blueshifts all lines. The quadratic effect is less pronounced for higher-$Z$ ions. The wavelength scale is {\it Chandra}'s wavelength scale (so models have been redshifted).
    }
    \label{fig:7}
\end{figure*}

The diagnostic plot is shown in 
\hyperref[fig:7]{Figure 7}, where the positions of the strongest transitions in H-like O and Ne as a function of magnetic field are apparent. We used the numerical calculations for an exact non-relativistic Hamiltonian for hydrogen given by \citet{Kravchenko1996}, scaling the energies and magnetic fields by $Z^2$. Note that the relativistic effects on the energies of the stationary states are of order $(\alpha Z)^2$ ($\alpha$ the fine structure constant), which are smaller than 0.53\% for $Z \leq 10$ and can be neglected in the present case ($\Delta\lambda \leq 0.05$ \AA\ at 15 \AA, or half a spectral bin). 
We applied a gravitational redshift of $z=0.29$ to these wavelengths. In our spectrum, the wavelength of the strongest \ion{O}{8} $n=1-3$ transition varies the most strongly as a function of $B$, while the longest-wavelength \ion{Ne}{10} line is relatively constant. Together, these features match the spectrum at $B = 2.9 \times 10^{10}$ G, and we very roughly estimate that the split between these features would no longer match if the wavelength difference were varied by more than $\approx 0.3$\AA; this corresponds to a change in $B$ of $\approx 10$\%. If we were to lower the field to $2.6 \times 10^{10}$ G, the gravitational redshift would decrease from $z=0.290$ to perhaps $z = 0.285$ (note how the centroid position between \ion{O}{8} $1s_0-3p_{0}$ and \ion{Ne}{10} $1s_0-2p_{-1}$ changes hardly at all as a function of magnetic field strength). Likewise, a similar small increase in 
$B$ would increase the redshift to $z = 0.295$.

The effective temperature is constrained mostly by the appearance of the observed $6-10$ \AA\ continuum. Too cool, and the spectrum does not rise high enough in that range; too hot, and it overproduces the short wavelengths $\leq 5$\AA. This limits the range to about
$T_{\rm eff} \approx 5.0 -5.5 \times 10^6$ K, with the chosen value slightly anticorrelated with the abundances of Ne, Mg, Si, and S (increasing the heavy element opacity tends to increase the flux well above the strongest absorption edge, pushing up the apparent 'Wien'-like tail of the spectrum).

The abundances of Ne and Si with respect to O are free to vary by about $\sim 30$\%, those of S and Ar (constrained by the $3-6$ \AA\ continuum) probably by a factor 2, and those of Mg, Ca, and Fe by factors of a few. At higher abundances, Fe starts producing noticeable opacity due to numerous discrete transitions in the Fe L shell ions in the $7-15$ \AA\ band.

We did not experiment much with the value of the surface gravity, and mostly kept it at $\log g = 14.6$. In principle, the width of the collision-broadened absorption lines depends almost linearly on the electron density in the X-ray photosphere, and therefore almost linearly on $g$. The treatment of the line profiles in our models is based on an explicit model for the atomic structure that explicitly includes the magnetic field, and it takes into account the interaction of the free electrons with the magnetic field as well \citep{Gomez2023,Gomez2024a,Gomez2024b}. Based on changes in the continuum at the shortest wavelengths, we very roughly estimate that $\log g$ is free to vary by $0.3$ or so. A more accurate measurement awaits more detailed modeling of the pressure broadening, and higher signal to noise data.

Finally, we comment on the fact that we are seeing the superposition of two hot regions on the stellar surface with slightly different characteristics. For instance, \citet{alford2022} determined that the broad-band variation of the spectrum as a function of spin phase could be understood by assuming that the two blackbodies in their analysis are approximately in antiphase. We are here seeing the spectrum of the sum of the two regions (note that unfortunately, due to a wiring issue with the HRC-S we cannot time-resolve our spectrum on the 0.112 sec spin period). To very rough approximation, keeping the other variables constant, the average of two atmospheres models at slightly different effective temperatures simply looks like the spectrum at the average effective temperature. With the range of variation in $T_{\rm eff}$ quoted above, we certainly are in that regime. Note that the relative sizes of the two spots do not really affect our analysis, and neither does uncertainty in the distance to the star: our analysis is entirely spectroscopic. And in any case, the gravitational redshift and acceleration of gravity should be identical for both regions, as will be, very likely, the set of abundances. Small variations in the effective temperature and magnetic field are of course possible. There may also be minor effects associated with the fact that we have simply used the flux spectra from our models (emergent intensity averaged over one
hemisphere), without allowing for the fact that the intensity varies across the surface, nor have we allowed for the effects of general relativistic light bending.
We expect these effects to mainly affect the details of line profiles, and the far Wien tail of the spectrum.

We briefly comment on the possibility that interstellar absorption lines may be present in the spectrum. We estimate the expected equivalent widths of the strongest transitions in H- and He like O and Ne, which should be present in the hot gas of the supernova remnant. The neutron star is seen to have moved away from a cloud of $\alpha$-element enriched gas that evidently marks the explosion site \citep{Mayer2022}. For our sight line to the neutron star, the absorbing remnant gas is shocked ISM, which presumably has roughly Solar abundances. Assuming a radius for the remnant of 10 pc, an average density of $n = 1$ cm$^{-3}$ \citep{Mayer2022}, abundances (by number) for O and Ne of O/H = $8.8 \times 10^{-4}$, Ne/H = $9 \times 10^{-5}$ \citep{Grevesse1998}, and assuming the lines are unsaturated, we find an equivalent width for \ion{O}{8} Ly$\alpha$ ($\lambda 18.97$ \AA) of 0.034 \AA, for \ion{Ne}{10} Ly$\alpha$ ($\lambda 12.13$ \AA) of $6 \times 10^{-3}$ \AA, and for \ion{Ne}{9} $w$ (the $n=1-2$ resonance line, $\lambda 13.45$ \AA) $0.011$ \AA; the corresponding resonance line in \ion{O}{7} is at 21.60 \AA, where the signal-to-noise is low. The lines will be unsaturated for the assumed ion column densities if the velocity widths of the lines are $\Delta v \gtrsim 500$ km s$^{-1}$, a width to be expected in an expanding young SNR. Of these, \ion{O}{8} Ly$\alpha$ may be detectable: a predicted $\sim 30\%$ depression in a 0.1 \AA\ bin (unresolved); the others are too small to be detectable in our data. There is a possible small feature, marginal at best, at a 
plausible wavelength: two bins at $\approx 18.3$ \AA\ are below the continuum (we assume the model stellar continuum shown in \hyperref[fig:6]{Figure 6}), with a depth of 28 counts against 128 continuum counts. That implies an equivalent width of 0.05 \AA, at $2.5\sigma$ significance. The lines would be blueshifted by $\sim 11,000$ km $^{-1}$ and possibly broadened by $\sim 1500$ km s$^{-1}$. Note that absorption lines due to the neutral intervening ISM are not detectable in our data; the shortest wavelength strong line would be \ion{O}{1} at 23.5 \AA\ \citep{paerels2001}.

\section{Conclusions} \label{sec:conclusion}

We have presented evidence for photospheric atomic absorption features in the X-ray spectrum of the neutron star RX J$0822-4300$ in the supernova remnant Puppis A. The spectrum shows absorption lines and edges consistent with a hot, high gravity atmosphere composed mainly of oxygen and neon, with small amounts of the heavier elements. We see the splitting and blueshifting of lines due to a strong magnetic field. The spectroscopically inferred field strength is consistent with the value measured from the star's spin and spindown rate. We determine a gravitational redshift of $z = 0.285-0.295$. 

In view of our finding that the X-ray photosphere is composed mainly of oxygen and neon, it is interesting that a recent image of the entire supernova remnant with {\it eROSITA} 
shows clouds of supernova ejecta of enhanced O, Ne, Si, S, and even Fe abundance at the explosion site \citep{Mayer2022}.

We compare our flux measurement against earlier measurements. From the counts in the LETGS spectrum we determine $\Omega \equiv (R/D)^2 = 9.92 \times 10^{-34}$ counts photon$^{-1}$, with $R$ the stellar radius [or rather $R^2$ equals the equivalent radiating surface area $A$, divided by $\pi$] and $D$ the distance. The measured bolometric luminosity of RX J$0822-4300$ is $L = \Omega \cdot 4\pi D^2 \sigma T_{\rm eff}^4 = 9.0 \times 10^{33}$ erg s$^{-1}$, for $T_{\rm eff} = 5.3 \times 10^6$ K, $\Omega = 9.9 \times 10^{-34}$. and $D = 1.3$ kpc. This puts the star on the cooling curve in \citet{Gusakov2004} (their Figure 1) at $\log [T/K]^{\infty} = 6.28$ if we assume $R = 10$ km, almost exactly where the {\it ROSAT} PSPC + {\it ASCA} GIS flux measurement given by \citet{Zavlin1999} places it (their bolometric flux, assuming a pure H atmosphere with a $10^{12-13}$ G field, when correcting to $D = 1.3$ kpc, is
$L_{\rm bol} = 0.7-1.1 \times 10^{34}$ erg s$^{-1}$).

\begin{figure*}
    \centering
    \includegraphics[scale=0.5]{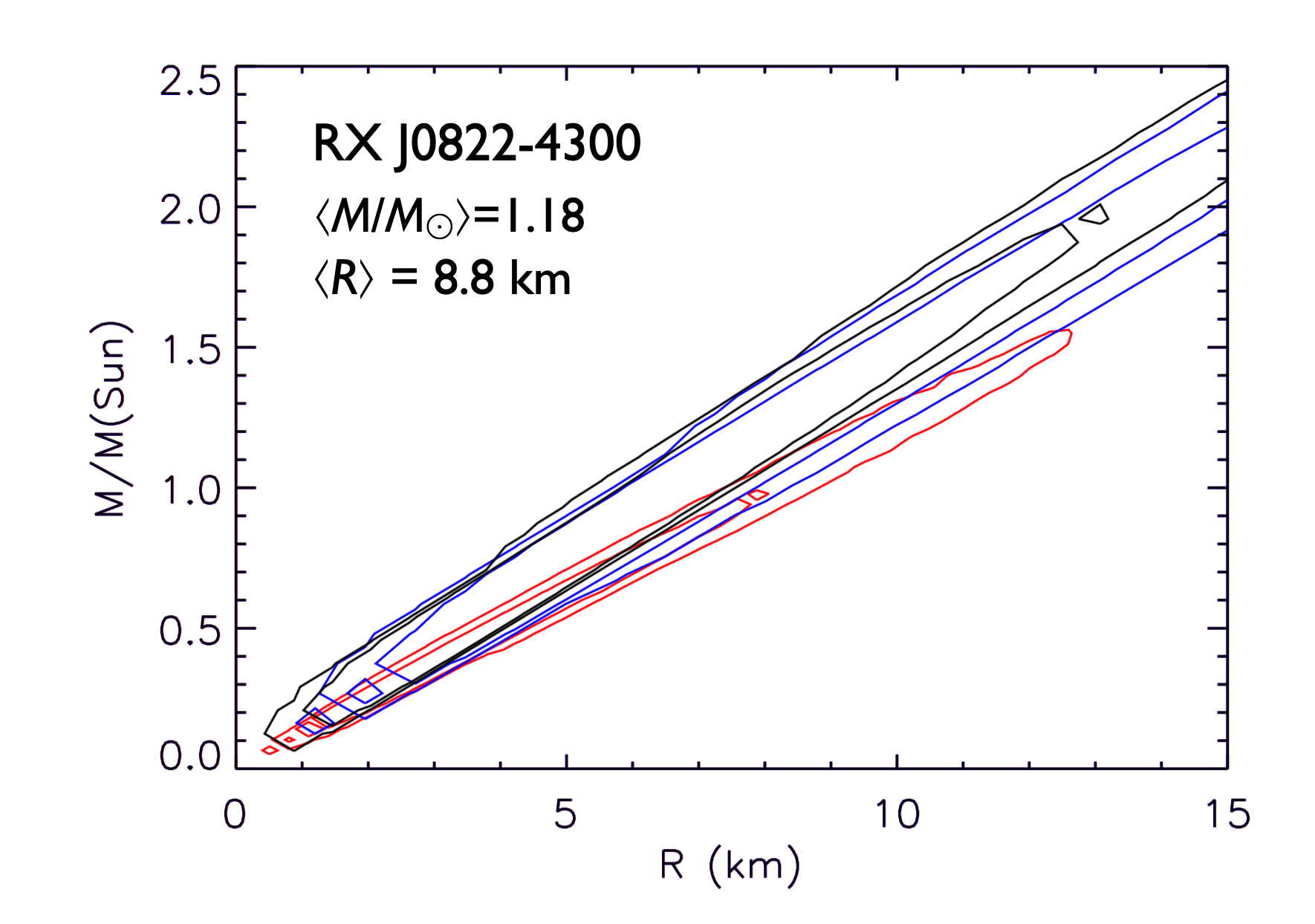}
    \caption{Probability density in the neutron star radius-mass plane, constructed assuming gravitational redshift $z = 0.29 \pm 0.02$, acceleration of gravity $\log g = 14.4 \pm 0.3$, with Gaussian distributions in $z$ and $\log g$ ({\it black contours}; the inner contour contains approximately 68\% of the probability, the outer contour approximately 90\%). The red contours are for the same assumptions, except the central value of the gravity is $\log g = 14.6$; the blue contours assume a central value of $\log g = 14.2$. 
    }
    \label{fig:8}
\end{figure*}

\begin{figure*}
    \centering
    \includegraphics[scale=0.45]{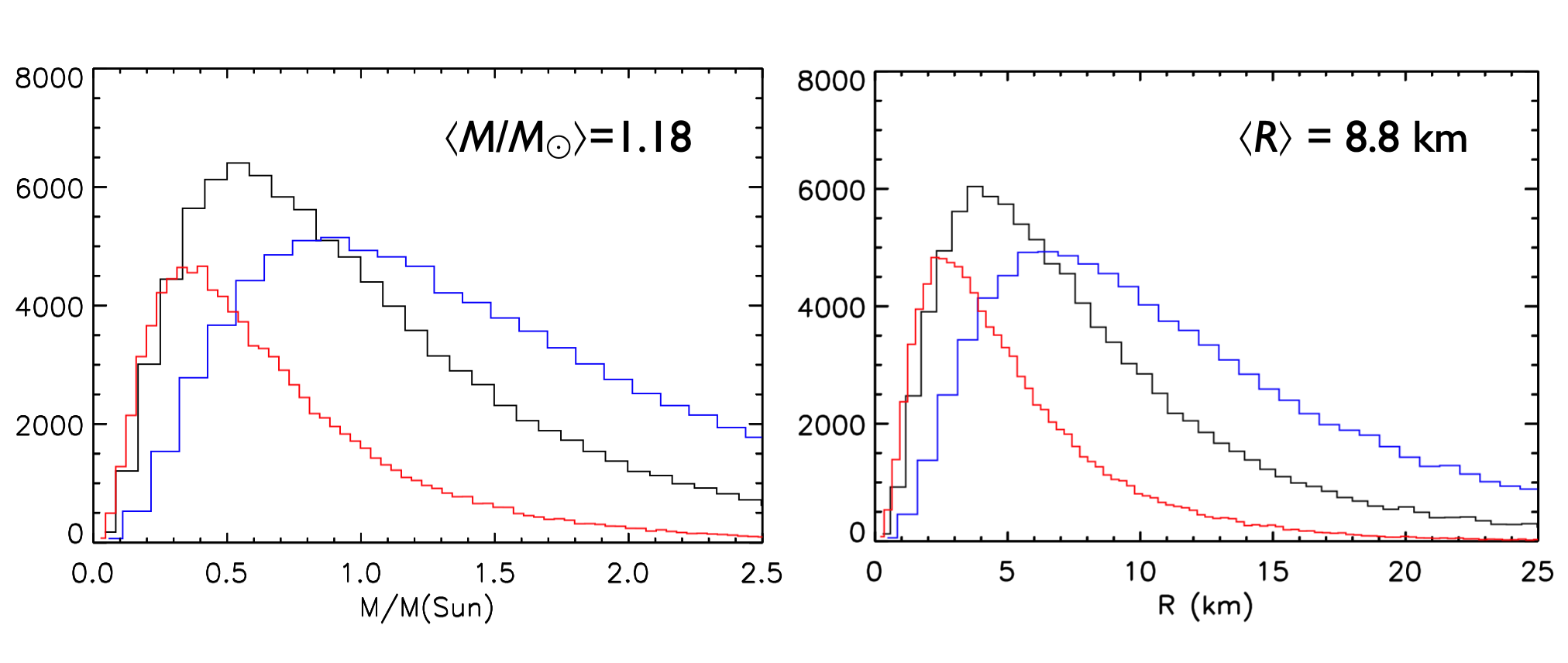}
    \caption{Projections of the probability onto the mass- and radius axes. The black, red, and blue histograms correspond to assumed central values of the gravity of $\log g = 14.4, 14.6,$ and $14.2$
    }
    \label{fig:9}
\end{figure*}

Finally: since we have both a measurement of the gravitational redshift and a constraint on the acceleration of gravity at the surface, we can in principle do a spectroscopic determination of the mass and the radius of the star. If we assume Gaussian probability distributions for the true values of the logarithm of the surface acceleration of gravity, and of the gravitational redshift, with central values and dispersions of $\log g = 14.4 \pm 0.3$ and $z = 0.29 \pm 0.02$, and we generate $10^5$ random realizations, we obtain the 
density plot in the radius-mass plane shown in \hyperref[fig:8]{Figure 8}; the density projected onto the mass and radius axes is shown in \hyperref[fig:9]{Figure 9}. The weighted average is at $M/M_{\odot} = 1.19$ and $R = 8.8$ km. We have chosen to loosen the constraints on the redshift from $\Delta z = 0.005$ to $0.02$ for this plot, to be conservative. The (approximate) 68\% and 90\% confidence areas show a wide range in both parameters, but with a strong correlation. This reflects the fact that we have a relatively precise measurement of the gravitational redshift, and only a rough constraint on the surface gravity. If this latter constraint can be improved, the error regions will shrink, primarily along the long axis in \hyperref[fig:8]{Figure 8}. This may be possible when we refine the 
modeling of the pressure broadening. 

As they stand, our mass and radius constraints are consistent with existing measurements and expectations, though perhaps the radius tends towards somewhat smaller values at the canonical mass value of $M/M_{\odot} = 1.4$ than have been obtained from observations of accreting millisecond pulsars. 

\begin{acknowledgments}
We gratefully acknowledge discussions with Brad Wargelin, who helped us decide which detector to use (satisfying the opposing demands of photon count versus background suppression). An earlier plan, to use the ACIS-S detector to suppress the nebular background by matching CCD-determined photon energies to grating dispersion angles, would, at the same count, have lowered the red line in \hyperref[fig:4]{Figure 4}, but it would by now require very long exposures, if feasible at all.
We also would like to express our gratitude to Harvey Tananbaum and Pat Slane for very helpful and detailed comments and criticism. We thank the referees for a critical reading of our manuscript.

J. G., F. P., and E. V. G. acknowledge support from NASA under grant number SAO GO0-21055X.
T.A.G. acknowledges support from the George Ellery Hale Post-Doctoral Fellowship at the University of Colorado and from the United States Department of Energy under grant DR-SC0010623 and the Wootton Center for Astrophysical Plasma Properties under the United States Department of Energy collaborative agreement DE-NA0003843.

This paper employs a list of Chandra datasets, obtained by the {\it Chandra} X-ray Observatory, contained in the Chandra Data Collection (CDC) 550~\dataset[doi:10.25574/cdc.550]{https://doi.org/10.25574/cdc.550}

\end{acknowledgments}

\vspace{5mm}
\facilities{{\it Chandra} X-ray Observatory}

\software{Ciao}

\newpage 

\bibliography{RXJ0822-4300refs}{}

\bibliographystyle{aasjournal}

\end{document}